\documentclass[prl,aps,floatfix,amsmath,amssymb,superscriptaddress,preprint]{revtex4-1}
\usepackage{latexsym}
\usepackage{graphicx}
\usepackage{tabularx}
\usepackage{times}
\usepackage[usenames,dvipsnames]{color}
\usepackage{amsmath}
\usepackage{amssymb}   
\usepackage{ulem} 
\usepackage{dcolumn}
\usepackage{latexsym,amsmath,amssymb,bm,euscript}
\usepackage{lineno}
\usepackage[
	colorlinks=true,
	urlcolor=BlueViolet,
 citecolor=Maroon,
	plainpages=false,
 	pdfpagelabels,
 	bookmarksnumbered
 	]{hyperref}

\def\mL{\mathcal{L}}
\def\mv{{\bf v}_\lambda}

\definecolor{gray}{rgb}{0.4,.4,0.4}
\definecolor{purple}{rgb}{0.6,.0,0.6}
\definecolor{darkgreen}{rgb}{0.0,.6,0.}

\def\beq{\begin{equation}}
\def\eeq{\end{equation}}

\begin{document}

\title{Species survival and scaling laws in hostile and disordered environments}

\author{Rodrigo P. Rocha}\email[]{rodrigo.rocha@ufsc.br}

\affiliation{Departamento de F\'\i sica, Universidade Federal de Santa Catarina, 
88040-900, Florian\' opolis-SC, Brazil.}
\affiliation{Dipartimento di Fisica e Astronomia, Universit\`a di Padova, CNISM and INFN, via Marzolo 8, I-35131 Padova, Italy.}

\author{Wagner Figueiredo}
\affiliation{Departamento de F\'\i sica, Universidade Federal de Santa Catarina, 
88040-900, Florian\' opolis-SC, Brazil.}

\author{Samir Suweis}
\affiliation{Dipartimento di Fisica e Astronomia, Universit\`a di Padova, CNISM and INFN, via Marzolo 8, I-35131 Padova, Italy.}

\author{Amos Maritan}
\affiliation{Dipartimento di Fisica e Astronomia, Universit\`a di Padova, CNISM and INFN, via Marzolo 8, I-35131 Padova, Italy.}

\date{\today}

\begin{abstract}
 In this work we study the likelihood of survival of single-species in the context of hostile and disordered environments. Population dynamics in this environment, as modeled by the Fisher equation, is characterized by negative average growth rate, except in some random spatially distributed patches that may support life. In particular, we are interested in the phase diagram of the survival probability and in the critical size problem, i.e., the minimum patch size required for surviving in the long time dynamics. We propose a measure for the critical patch size as being proportional to the participation ratio (PR) of the eigenvector corresponding to the largest eigenvalue of the linearized Fisher dynamics. We obtain the (extinction-survival) phase diagram and the probability distribution function (PDF) of the critical patch sizes for two topologies, namely, the one-dimensional system and the fractal Peano basin. We show that both topologies share the same qualitative features, but the fractal topology 
requires higher spatial fluctuations to guarantee species survival. We perform a finite-size scaling and we obtain the associated scaling exponents. In addition, we show that the PDF of the critical patch sizes has an universal shape for the 1D case in terms of the model parameters (diffusion, growth rate, etc.). In contrast, the diffusion coefficient has a drastic effect on the PDF of the critical patch sizes of the fractal Peano basin, and it does not obey the same scaling law of the 1D case.  
\end{abstract}


\maketitle

\section{I. Introduction}
Finding the conditions for extinction or survival of a species in a given environment is a very important challenge attracting considerable attention of the natural science community \cite{May1976,Brock1999,Chesson2000a,Chave2002,Murray2003}. In particular, many studies have highlighted the important role that spatial connectivity \cite{Tilman,Kneitel2004,Bertuzzo2011,Carrara2012} and environmental heterogeneity \cite{Chesson2000b,Grilli2013,Azaele2016} play in deterring the species lifetime. In this work we tackle two research questions related to population dynamics in the context of hostile and disordered environments. The first one is related to the conditions that lead to extinction as well as survival of species, i.e., the survival probability $P_s$. The second question is to find the minimum critical patch size needed to ensure survival of species in the long time dynamics, i.e., the critical patch size $L_c$ (see Refs. \cite{Okubo2001,Ryabov2008} for a review). Among many important applications, 
these concepts have been applied to design national parks, natural reserves \cite{Cantrell1998,May1976}, protection of endangered species \cite{Brock1999} etc.

In continuous deterministic (mean-field) population dynamics models, the spatiotemporal evolution of the population density is typically described by a reaction-diffusion equation with a logistic growth. The very successful Fisher equation is a particular case \cite{Fisher1937,Murray2003}, where the logistic growth assumes the quadratic form:
\begin{equation}
\frac{\partial \rho(x,t)}{\partial t}=D\nabla^2 \rho(x,t)+\mu\rho(x,t)
-b\rho^2(x,t),
\label{continuous_fisher}
\end{equation}
where $\rho(x,t)$ is the population density, $D$ is the diffusion coefficient (which accounts for the effect of migration), $\mu$ is the growth rate and $b$ is the death rate. Eq. \eqref{continuous_fisher} has been employed to study a wide variety of systems, including dynamics of bacteria \cite{Nadav2000,Neicu2000,Kenkre2003}, epidemiology \cite{Murray2003}, chemical kinetics \cite{Douglas2007}, to name just a few of them.

The critical size problem related to equation \eqref{continuous_fisher} has a long history \cite{Okubo2001,Ryabov2008,Kierstead1953,Skellam1951}. In its simplest one-dimensional version, population undergoes a logistic growth ($\mu>0$) along a favorable patch of size $L$, while it is surrounded by a totally hostile environment with infinite death rate, i.e., when the population reach the habitat boundaries, it is absorbed, killed or removed instantaneously. This model is known as the KiSS size model, and was independently introduced by Kierstead and Slobodkin \cite{Kierstead1953} and Skellam \cite{Skellam1951}. The critical patch, $L_c$, may be obtained linearizing \eqref{continuous_fisher} around $\mathbf{\rho=0}$ and assuming Dirichlet boundary conditions, i.e., $\rho(L/2,t)=\rho(-L/2,t)=0$. A straightforward calculation leads to
\begin{equation}
L_c=\pi\sqrt{D/\mu}.
\label{KiSS}
\end{equation}
The same functional dependence of $L_c$ on $D$ and $\mu$ holds for a two-dimensional system up to a multiplicative factor \cite{Kierstead1953,Skellam1951}. Therefore, the likelihood of species extinction is certain for $L<L_c$, while the species will survive with probability one if $L>L_c$.

However, the assumptions of infinite death rate outside the favorable patch as well as an uniform growth rate inside it (i.e., a homogeneous environment) are idealizations and much attention has been given to model more realistic settings. Extensions of the KiSS model have evolved in two main directions. On the one hand, many studies have focused on modeling the hostile environment employing different kinds of growth functions and boundary conditions, but still using the mean-field dynamics given by Eq. \eqref{continuous_fisher}. In this direction we may mention, for example, heterogeneous growth functions with gradual variation \cite{Kenkre2011}, finite mortality outside the favorable patch \cite{Nel2000}, and many others \cite{Kraenkel2010,Mendes2008,Okubo2001,Murray2003}.

On the other hand, the main motivation is to develop more realistic population dynamics models, where the mean-field description fails, for instance, due to the discrete nature of the population abundances (demographic fluctuations) \cite{Durrett1994,Azaele2016}. Several efforts have been done in this direction. One can also resort to different formalisms, like a master equation approach \cite{Escudero2004}, stochastic partial differential equations \cite{Mendes2010}, or employ a discrete particle model \cite{Berti2015}. However, in many cases, the corresponding growth rates are uniform (like in Refs. \cite{Escudero2004,Berti2015}), or a completely hostile environment outside the favorable patch is assumed (like in Ref. \cite{Mendes2010,Berti2015}). 

In fact, in many experimental conditions and in most of the real cases, the environment is neither static nor spatially constrained (i.e., oasis-desert picture): rather, it may fluctuate in space as, for example, captured by a random spatial disorder in the species dynamics \cite{yachi1999,kussell2005}. Since the critical patch is a function of the spatial random configuration of disorder the critical patch itself is a random variable whose probability distribution function (PDF) has never been calculated in all the extensions of the KiSS model mentioned above.

Our goal in this work is to determine the survival probability and the PDF of the critical patch sizes in a disordered environment. We shall employ the Fisher equation \eqref{continuous_fisher} and use random growth rates as a null model for environmental fluctuations.

More precisely, we shall address the following specific issues regarding the (random) critical patch size problem: $i)$ We start our analysis investigating the one-dimensional system with periodic boundary conditions. Following \cite{Nel98}, we define a survival criterion based on the largest eigenvalue of the matrix governing the linearized Fisher dynamics. We will then obtain the phase diagram of the survival probability and perform finite-size scaling to obtain scaling exponents. $ii)$ In what concerns the random critical patch size, we show that the participation ratio (PR) of the eigenvector corresponding to the largest eigenvalue of the linearized dynamics may be used to estimate the critical patch size. So far, this connection has not been employed to calculate $L_c$ in a systematic way, and this is an original contribution of this work. First, we apply this ansatz to show that the average value of the PR has the same functional dependence on the critical patch as the KiSS model. In addition, we will provide an explicit analytical expression for the probability distribution function of the critical patch sizes. We will show that such distribution has an universal functional shape in terms of the model parameters, and we will obtain its finite-size scaling properties. $iii)$ Once we have validated our ansatz to estimate $L_c$, we employ the same tool to study the random critical patch size problem in a more complex topology than the 1D system. As a particular example, we investigate the fractal Peano basin \cite{PeanoStructure1996,PeanoStructure1997}. Important applications of this fractal topology arise mainly in the context of river networks \cite{RinaldoBook}. There are many studies of the Fisher equation applied to this topology \cite{PeanoFrontWave2006,PeanoFrontWave2005,RinaldoBook}, but little attention has been given to the phase diagram of the survival probability and the critical patch size. We start by showing that the phase diagram has the same qualitative features of the one-dimensional case, although it is described by other scaling exponents. On the other hand, we show that the average value of the PR is no longer proportional to the square-root of the diffusion coefficient. The distribution of the critical patch sizes is more complex than in the linear case, and does not obey the same scaling properties.  

Our work is organized as follows: in section II we will present the survival criterion used to study extinction and survival of the species. This criterion is based on the largest eigenvalue of the matrix governing the linearized dynamics. Then we will define the survival probability in terms of the PDF of this largest eigenvalue. In section III we will show our numerical results for the one-dimensional system. First, we will address the survival probability and then we will discuss the random critical patch size. In section IV we show our results for the fractal Peano basin. A conclusion section closes the paper. In the appendix we derive an analytical expression for the PDF of the critical patch sizes.

\section{II. The criterion for surviving}
This section is dedicated to present the survival criterion used to study extinction and survival for a species population undergoing a Fisher dynamics in a disordered environment, following the work of Nelson and Shnerb \cite{Nel98}. 

The discrete version of Eq. \eqref{continuous_fisher} is
\begin{equation}
\frac{\partial \rho_i(t)}{\partial t}=\sum_{j=1}^{N}\mL_{ij}\rho_j(t)-b \rho_i^2(t), 
\label{fisher}
\end{equation}
where $\rho_i(t)$ is the population density at site $i$. We assume homogeneous initial conditions, i.e., $\rho_i(0)=\rho_0>0$ for all sites $i=1,\cdots,N$ of the discrete network. The Liouville operator, $\mL$, governs the linearized dynamics around $\mathbf{\rho=0}$. The latter is a symmetric random matrix with elements given by
\begin{equation}\label{L}
\mL_{ij}=\begin{cases} (D/\ell_0^2)M_{ij} & \text{if $i\ne j$}, \\
\mu_i-(D/\ell_0^2)\sum_k M_{ik} & \text{if $i = j$}, \end{cases} 
\end{equation}
where $\ell_0$ is the lattice parameter and $\mu_i$ is the random growth rate at site $i$. We assume that $\mu_i=-a+U_i$, where $a$ is a positive constant indicating different levels of hostility and $U_i$ is an independent and identically distributed random variable drawn from the uniform distribution in the interval $[-\Delta,\Delta ]$, where $\Delta$ is the parameter characterizing the strength of the environmental fluctuations on the growth rate. 

The adjacency matrix $M$ entering in \eqref{L} accounts for network topology. Its matrix elements are $M_{ij}=1$ if sites $i$ and $j$ are connected to each other and $M_{ij}=0$ otherwise. We expand $\rho_i(t)$ in a complete set of eigenvectors $\mv$ of $\mL$, i.e.
\begin{equation}
\rho_i(t)=\sum_{\lambda} c_\lambda(t)(\mv)_i,
\label{guess}
\end{equation}
where $(\mv)_i$ is the i-th component of $\mv$, and the sum is performed over all eigenvalues of $\mL$. We assume ortonormalized eigenvectors, i.e., $\sum_i ({\bf v}_{\lambda})_{i} ({\bf v}_{\lambda'})_{i}=\delta_{\lambda,\lambda'}$, where $\delta_{\lambda,\lambda'}$ is the Kronecker delta. In this way the set of coupled dynamical equations becomes, 
\begin{equation}
\frac{d c_\lambda(t)}{dt}=\lambda c_\lambda(t)-\sum_{\lambda',\lambda''}w_{\lambda,\lambda'\lambda''}
c_{\lambda'}(t)c_{\lambda''}(t),
\label{dynamical_c}
\end{equation}
where the coupling coefficients $w_{\lambda,\lambda'\lambda''}$ are defined as
\begin{equation}
w_{\lambda,\lambda'\lambda''}=b\sum_i (\mv)_i({\bf v}_{\lambda'})_i({\bf v}_{\lambda''})_i.
\label{coupling_c}
\end{equation}
Up to this point the analytical treatment is exact, but calculation of \eqref{dynamical_c} is challenging. We will explore an analogy, in the imaginary time, between $\mL$ and the Anderson localization Hamiltonian $H$ \cite{Nel98}, to obtain an approximate solution of \eqref{dynamical_c}. Indeed, we know \cite{And58,Kramer93} that for a 1D disordered Hamiltonian, like the one given by Eq. \eqref{L}, the eigenvectors corresponding to the few positive eigenvalues are localized and, to a first approximation, non-overlapping. Therefore the coupling coefficients $w_{\lambda,\lambda'\lambda''}$ are almost negligible unless $\lambda'=\lambda''=\lambda$. This approximation decouples \eqref{dynamical_c} and we get the solution  
\begin{equation}
c_\lambda(t)=\frac{c_\lambda(0)e^{\lambda t}}{1+c_\lambda(0)(w_\lambda/\lambda)(e^{\lambda t}-1)},
\label{approximated_a}
\end{equation}
where $w_\lambda \equiv w_{\lambda,\lambda\lambda}= b\sum_i [(\mv)_i]^3$. Upon replacing \eqref{approximated_a} in \eqref{guess} and taking the limit $t\to \infty$ we obtain the steady state concentration
\begin{equation}
  \rho^{\star}_i=\sum_{\lambda > 0} (\lambda/w_\lambda)(\mv)_i,
\label{c*}
\end{equation}
where the sum is performed only over the positive eigenvalues of $\mL$. The stationary state, Eq. \eqref{c*}, is independent of the initial conditions as far as the  $c_\lambda(0)$'s are all different from zero. The stationary value of the total population simply becomes
\begin{equation}\label{K}
K^\star=\sum_{i,\lambda > 0} (\lambda/w_\lambda) (\mv)_i. 
\end{equation}
Equations \eqref{c*} and \eqref{K} constitute the main results of this section. Indeed, we observe that $\rho_i^{\star}$ and $K^{\star}$ are proportional to a sum over localized eigenvectors $\mv$ ($\lambda>0$) and, in the limit of few positive eigenvalues, population will survive and remain localized around small patches in space. Therefore, analytical condition on the largest eigenvalue $\lambda_1$ of $\mL$ allow us to predict species survival ($\lambda_1>0$), or extinction ($\lambda_1<0$), without needing to integrate the entire Fisher equation \eqref{continuous_fisher}.

Notice that, although the presence of quenched random growth rates in Eq. \eqref{fisher}, the population density $\rho_i(t)$ still evolves in a deterministic way. In addition, the largest eigenvalue of $\mL$ is a variable that depends on the particular system configuration, i.e., $a, D, N, \Delta$ and $\vec U=\{U_i\}$. In this case the conditional probability, $p(\lambda_1|\vec U)$, simply becomes $p(\lambda_1|\vec U)=\delta(\lambda_1-\lambda_1(\vec U))$. Therefore, the probability distribution function (PDF) of $\lambda_1$ is obtained averaging the conditional probability, $p(\lambda_1|\vec U)$, over the distribution, $p(\vec U)$, of $\{U_i\}$, i.e.,
\begin{equation}\label{lambda1PDF}
p(\lambda_1)=\int p(\lambda_1|\vec U) p(\vec U)d\vec U.
\end{equation}
Unfortunately, one can not solve the above equation analytically and one must resort to numerical calculations to obtain  $p(\lambda_1)$.

In order to quantify the approach to extinction (survival), we will define a survival probability, $P_s \in [0,1]$, according to the following expression:
\begin{eqnarray}
P_s = \int_0^\infty p(\lambda_1) d\lambda_1.
\label{probability}
\end{eqnarray}
The survival probability is the central quantity of our analysis. Accordingly, we define the following phases:
\begin{eqnarray}
 &P_s=0,& \qquad \text{extinction phase},\\
 &0<P_s<1,&\qquad \text{coexistence phase},\\
 &P_s=1,&\qquad \text{survival phase}. 
\end{eqnarray}
We will use this tool to address the phase diagram of the survival probability.

On the other hand, we observe that the spatial extent of the (localized) eigenvectors of $\mL$ reflects the spatial extent of $L_c$. The participation ratio (PR) is a standard quantity used to study eigenvector localization \cite{IPR1970}. It is defined by:
\begin{equation}
r({\bf v})=\frac{(\sum_i |v_i|^2)^2}{\sum_i |v_i|^4},
\label{PR}
\end{equation}
and is roughly equal to the number of sites where the eigenvector has a significant weight. In the uniform case ($\Delta=0$),  $v_i\sim 1/\sqrt{N}$ and $r\sim N$. In the limit of strong disorder, $r\sim 1$, and the eigenvector is localized over a single site. It is important that the definition of the participation ratio is independent of a particular eigenvector normalization, i.e., $r({\bf v})=r({\bf v}^{\prime})$, where ${\bf v}^{\prime}=c{\bf v}$ and $c$ is a (real or complex) constant. We will show, in the next section, that the participation ratio of the eigenvector associated to $\lambda_1$  may be used to estimate the critical patch size $L_c$.

\section{III. The One-Dimensional case}
In this section we show our numerical results for $P_s$ and $L_c$, for the one-dimensional system with periodic boundary conditions. We computed the first five largest eigenvalues of $\mL$ and the corresponding eigenvectors using ARPACK routines \cite{ARPACK}. Using this package we were able to consider systems up to order $N\sim 10^4$. For each value of the disorder parameter $\Delta$ we perform averages over approximately $2\times 10^4$ samples. We present our results in some suitable unit of time and we assume, for convenience, that $\ell_0=1$.

\subsection{The survival probability}
We now focus on the impact of spatially random growth rates on the survival probability $P_s$. Although the critical patch size is closely related to the survival probability, we shall dedicate a specific subsection to discuss this relation later.

\begin{figure}
\includegraphics[width=\columnwidth]{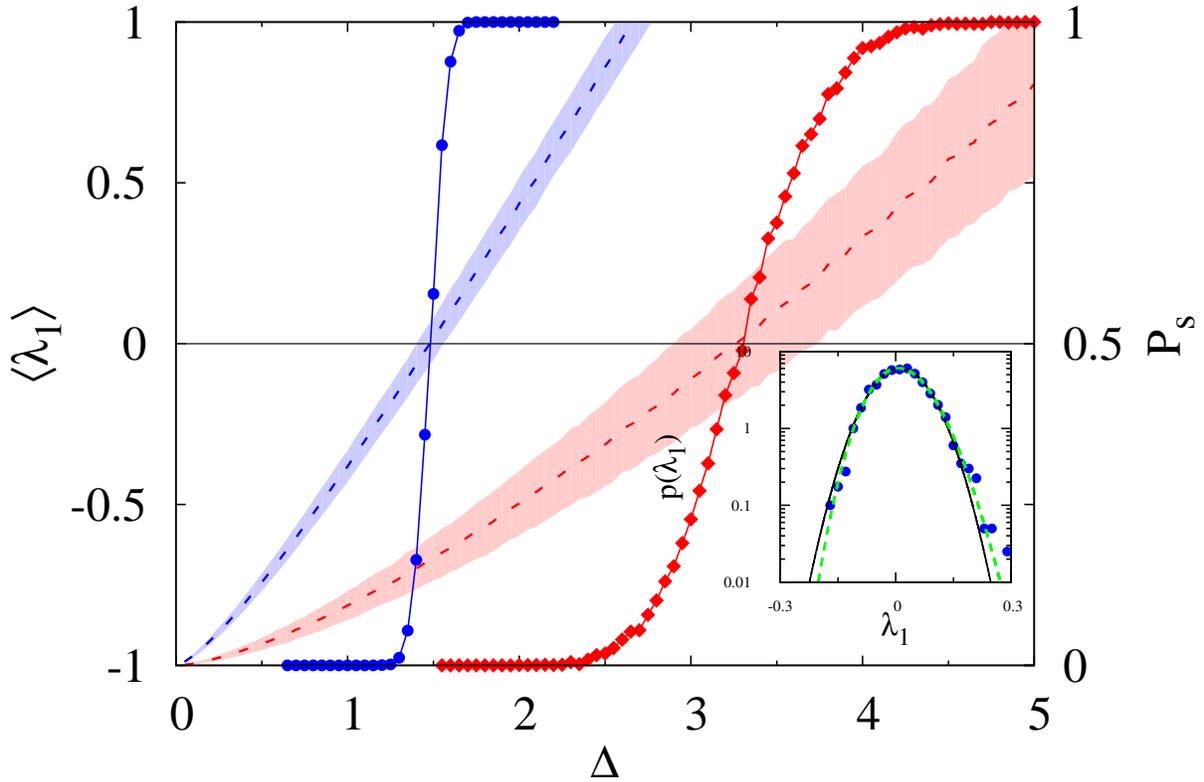}
\caption{Left vertical axis: Average of $\lambda_1$ as a function of disorder strength $\Delta$ (dashed line). In the shaded region we plot the standard deviation. Right vertical axis: Survival probability $P_s$ as a function of disorder strength $\Delta$ (dotted lines). The system size is $N=1000$, and the diffusion coefficients considered are $D=1$ (blue) and $D=30$ (red). In the inset we show in a semi-log scale $p(\lambda_1)$ for $\Delta=1.5$ and $D=1$. The peak of the distribution is well centered around $\langle \lambda_1 \rangle$. The green dashed line corresponds to a Tracy-Widom fit (four fitting parameters, see Ref. \cite{chiani2014}), while the black line corresponds to a Gaussian fit (two fitting parameters).}
\label{fig1}
\end{figure}

In Fig. \ref{fig1} we show the survival probability for a system with $1000$ sites and $a=1$. In the same plot, we show the average of the largest eigenvalue (left vertical axis). The uniform environment is simply recovered for $\Delta=0$. In this case the eigenvalue problem for $\mL$ can be exactly solved \cite{Economou}, resulting in $\lambda_1=-a$. Therefore, the survival probability is zero for $a>0$ and the species becomes extinct.

Now consider the disordered environment ($\Delta\ne0$). Complete extinction occurs with probability 1 for $\Delta <a$. Above this threshold $P_s$ is a smooth function of $\Delta$ and there is a phase supporting coexistence of both extinction and survival of species. 

Diffusivity has a negative impact over species survival, i.e., the  value of $\Delta$ necessary to keep $P_s$ at value $1/2$ increases with $D$. Furthermore, diffusivity increases the variance of $\lambda_1$ (the standard deviation of $ \lambda_1$ is represented by the shaded region of Fig. \ref{fig1}), which causes the broadening of the coexistence phase ($0<P_s<1$). This can be seen in Fig. \ref{fig1} for the values $D=1$ and $D=30$.

In order to characterize $P_s$, we need to compute the probability distribution function of the largest eigenvalue of $\mL$, $p(\lambda_1)$. However, results from the random matrix theory indicate that for various classes of random Hermitian matrices, the probability distribution of the normalized largest eigenvalue is universal \cite{tracy2002}, i.e. $p(\lambda_1)$ has an universal functional shape known as Tracy-Widom distribution \cite{tracy2002}. Indeed, it is known that a very good approximation to the Tracy-Widom distribution is the Gamma distribution \cite{chiani2014}, that in turn, for a large shape parameter, converges to a Gaussian distribution.

The random matrix \eqref{L} governing the linearized dynamics is Hermitian with random elements only in the diagonal. However, we find numerically that even our class of matrices seem to obey to Tracy-Widom distribution. We fitted, for a particular set of parameters, $p(\lambda_1)$ to a Gaussian and a Tracy-Widom distribution. Our main results are shown in the inset of Fig. \ref{fig1}. We observe that $p(\lambda_1)$ is slightly asymmetric. Moreover, the Gaussian fit deviate little from the Tracy-Widom fit. Given this small difference, we will approximate $p(\lambda_1)$ to a Gaussian distribution.

Now we investigate the phase diagram. Without loss of generality, we define the critical disorder strength, $\Delta_c(N,D,a)$, according to the implicit expression $\langle \lambda_1\rangle(\Delta_c,N,D,a)=0$. In this way we have  $P_s(\Delta,N,D,a)\to1/2$ for $\Delta\to \Delta_c$, which is consistent with a Gaussian approximation for $p(\lambda_1)$. Therefore, $\Delta_c$ reflects a sort of critical extinction-survival transition. To extract $\Delta_c$ from our numerical simulations we fit $\langle \lambda_1\rangle$ to a polynomial function of degree two. The fitting is done in a small interval in $\Delta$  around $\langle \lambda_1\rangle=0$ where the chi-squared becomes $< 10^{-5}$. In this way, $\Delta_c$ is simply obtained by solving the quadratic equation $\langle \lambda_1 \rangle=0$.

\begin{figure*}
\includegraphics[width=\textwidth]{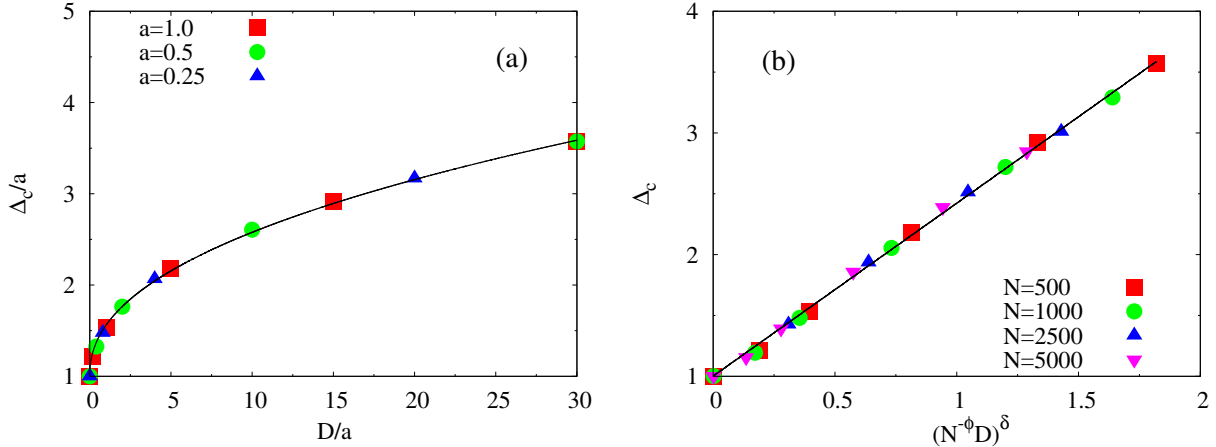}
\caption{Panel (a): Critical disorder strength $\Delta_c/a$ vs. $D/a$ for a one-dimensional system with $N=500$. The growth rates considered are shown in the figure. The solid line is a fit according to expression \eqref{powerlaw}. The reduced chi-squared is $< 5\times 10^{-4}$. Panel (b): We linearize $\Delta_c$ using the exponent $\delta=0.45$ and show the collapse for a scaling function in the form $f=N^{-\phi\delta}$, with $\phi=1/3$ being the scaling exponent. The growth rate considered is $a=1$.}
\label{fig2}
\end{figure*}

From a practical point of view, we have to determine $\Delta_c$ in a three-dimensional parameter space ($a$, $D$, and $N$). To face this problem we used the following methodology. First, we fixed a certain arbitrary value of $N$. Then we analyzed the dependence of $\Delta_c/a$ on $D/a$ (indeed in Eq. \eqref{fisher} one can absorb $a$ in a redefinition of time and the effective parameters become $\Delta/a$, $D/a$ and $b/a$). The main results of this analysis may be seen in Fig. \ref{fig2} (a).  In the limiting case of a vanishingly diffusivity, $D\to 0$, the critical strength approaches $a$ from above ($\Delta_c\to a$), given that the system size is large enough ($N \gg1$). We find that $\Delta_c$ has a power-law dependency, $\Delta_c/a = f(N)(D/a)^{\delta}+1$, where $f$ is a function of $N$ and $\delta$ is an universal exponent. The solid line in Fig. \ref{fig2} (a) is a fit according to this expression (the chi-squared is less than $\sim 5\times 10^{-4}$). To verify universality of $\delta$, we repeated this same procedure for different values of $N$. We find from the fitting analysis a small fluctuation of $\delta$ less than $5\%$, hence, $\delta=0.45\pm 0.02$.

The dependence of $\Delta_c$ on $N$ was obtained using a different technique. In this case we employed a finite-size scaling analysis \cite{Cardy1988}. Fig. \ref{fig2} (b) shows a quite good collapse for a scaling function in the form $f(N)=N^{-\phi\delta}$, where $\phi=1/3$. The actual value of $\phi$ was guessed from the quality of the collapse.

Finally, we can write out explicitly the critical disorder strength as
\begin{equation}
\Delta_c(N,D,a)=ca^{1-\delta}(N^{-\phi}D)^{\delta}+a,\qquad (N\gg 1),
\label{powerlaw}
\end{equation}
where $\phi=1/3$, $\delta=0.45\pm 0.02$, and $c=1.42\pm 0.02$. Equation \eqref{powerlaw} constitutes our first fundamental result. In a disordered environment we are able to predict how different parameters ($D$, $N$ and $a$) affects the critical disorder strength needed to make $P_s=1/2$. 

\subsection{The random critical patch sizes}
As explained in the introduction, in the case of an uniform environment (KiSS model), the critical patch size is known \cite{Kierstead1953,Skellam1951}, and $L_c$ is proportional to the square-root of the diffusion coefficient (see Eq. \eqref{KiSS}). The critical patch size for a species population undergoing a Fisher dynamics in a fluctuating spatially random environment is not known, and to find it numerically is a difficult and time consuming task.

The fundamental difference from the uniform KiSS model and its extensions \cite{Kenkre2011,Nel2000,Kraenkel2010,Escudero2004,Mendes2010,Berti2015} is the random nature of the critical patch size. Here we use $F(L_c)$ to represent the PDF of the critical patch sizes and $\langle L_c \rangle$ to represent its ensemble average.

We address the critical patch size using the ansatz for $L_c$ as given by Eq. \eqref{PR}:
\begin{equation}
L_c/\ell_0\approx r({\bf v}_{\lambda_1}),
\label{criticalpatch}
\end{equation}
where $\ell_0$ is the lattice parameter (fixed to $\ell_0=1$ for convenience) and $\lambda_1$ is the largest eigenvalue of $\mL$. When $D=0$, the above ansatz is clearly satisfied: the critical size of the patches in order the species to survive is 1 (in each site when $(-a+U_i)>0$ is satisfied), that is exactly the value of the PR corresponding to the largest eigenvalue of $\mL$ (that for $D=0$ is a diagonal matrix). For one positive eigenvalue ($\lambda_1>0$), from Eq. \eqref{c*} we know that only sites where the species will survive correspond to those entries where ${\bf v}_{\lambda_1}$ is localized, and thus the ansatz is again verified. Finally in the case of few positive eigenvalues, we will have more than one patches where the species survives. From the Anderson localization (and assuming non-overlapping eigenvectors), we know that the critical (smallest) patch size $L_c$ corresponds to the PR of the eigenvector associated to the largest eigenvalue of $\mL$ \cite{Kramer93}. We note that our ansatz is supported by the intuitive meaning of the definition of the PR. Indeed, when the system is localized in a single site $r=1$ and $L_c=1$, while when it is fully delocalized $r=N$ and $L_c=N$. Therefore, our ansatz is a generalization to all intermediate cases. For example if $v_i=\exp{(-|i|/\xi)}$ one finds $r=\coth^2(1/\xi)/\coth(2/\xi) \approx 2\xi$ and so $L_c/\ell \approx \xi$ as intuitively expected.

\begin{figure*}
\includegraphics[width=\textwidth]{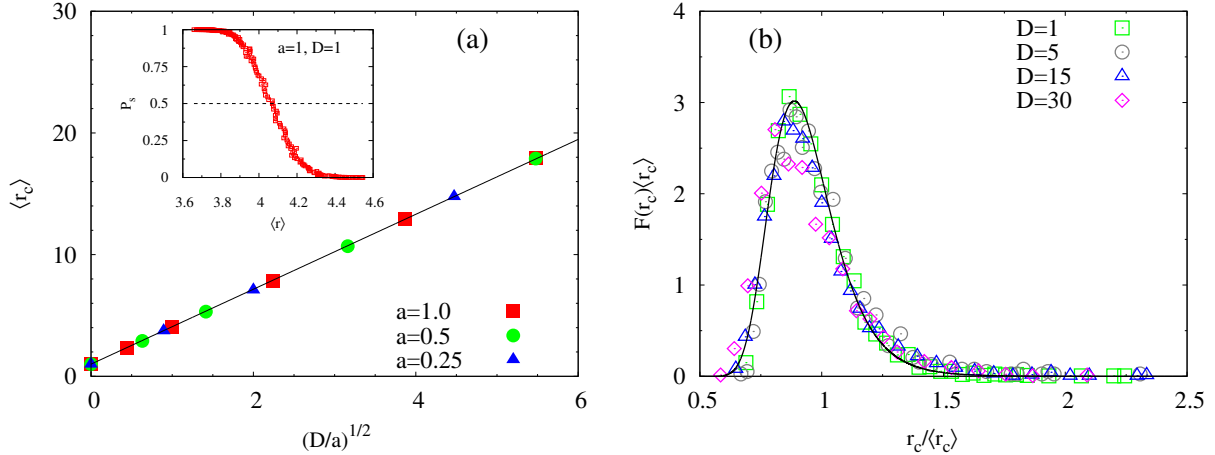}
\caption{Panel (a): Average of the participation ratio along the critical disorder strength (such that, $P_s\to 1/2$) against $\sqrt{D/a}$ for a 1D system. We consider three levels of hostility $a$ as shown in the legend. The solid line corresponds to expression \eqref{criticalL}. In the inset we show the relation between $P_s$ and $\langle r \rangle$. The parameters considered are shown in the figure. Panel (b): Collapse of the PDF of participation ratios according to expression \eqref{pL}. We fixed the parameters $N=1000$ and $a=1$. Notice that the small fluctuation observed is related to histograms (bin) problems.}
\label{fig3}
\end{figure*}

The way $\langle r \rangle$ is related to the survival probability $P_s$ is shown in the inset of Fig. \ref{fig3} (a), for a particular set of parameters (shown in the figure). The critical patch size is small for large values of $P_s$ (population can survive in small patches). In fact, $P_s$ increases with  $\Delta$ (see Fig. \ref{fig1}), meaning that environmental fluctuations favor species persistence. The same behavior has been reported in \cite{Mendes2010}. We now investigate the behavior of $\langle r_c \rangle$ and $F(r_c)$ along the critical disorder strength $\Delta_c$ (such that, $\langle \lambda_1 \rangle\to 0$ and $P_s\to 1/2$). In particular, we investigate their dependence on $D/a$ and $N$, the latter studied by means of finite-size scaling. In Figs. \ref{fig3} (a) and \ref{fig3} (b) we present the summary of our main numerical results.

The main box of the Fig \ref{fig3} (a) shows $\langle r_c \rangle$ as a function of $\sqrt{D/a}$ for three different levels of hostility, namely, $a=1$, $a=1/2$ and $a=1/4$. The perfect straight line observed leaves no doubt about the dependence of $\langle r_c \rangle$ on the square-root of $D/a$, exactly the same dependence of $L_c$ on $D/a$ in the KiSS model (see Eq. \eqref{KiSS}). We analyzed the scaling of $\langle r_c \rangle$ on $N$. As expected our results indicate a very weak and negligible dependence of $\langle r_c\rangle$ on $N$, with an exponent compatible with zero.

Finally, we can write out explicitly the critical patch size as
\begin{equation}
\langle L_c \rangle/\ell_0 \approx \langle r_c \rangle = C \Big(\frac{D}{a} \Big)^{\beta} +1,
\label{criticalL}
\end{equation}
where $\beta=1/2$ and $C=2.79\pm 0.01$. Observe that for $D\to 0$ we have $\langle L_c \rangle/\ell_0\to 1$, meaning localization of the population over a single site, which is the correct result.

To achieve a complete characterization of the problem, we investigated the PDF, $F$, of the participation ratios along the critical disorder strength. Finite-size scaling arguments suggest that:
\begin{equation}
 F(r_c)=\frac{1}{\langle r_c\rangle}p\Big(\frac{r_c}{\langle r_c \rangle}\Big),
\label{pL}
\end{equation}
where $p(x)$ has an universal shape in terms of the model parameters, and is given by
\begin{equation}
\label{fL}
p(x)= \frac{A}{x^2}\exp\Big\{-\frac{(\frac{2}{x}-\mu)^2}{2\sigma^2} \Big\}, \qquad x > 0,
\end{equation}
where $A$ is a normalization constant, $\mu$ and $\sigma$ are free parameters that can be fixed from a fitting procedure. A simple derivation of the expression \eqref{fL} is provided in the appendix. The solid line in Fig. \ref{fig3} (b) is a fit with Eq. \eqref{fL}: the best fit gives $\mu=0.48\pm 0.01$ and $\sigma=0.071\pm 0.005$.

Equations \eqref{criticalL}-\eqref{fL} constitute our second main result. We have a full characterization of the critical patch sizes as a function of the different parameters ($D$, $N$ and $a$): given a species in a hostile and disordered environment, we can determine in which patches the population will survive in the long time dynamics.

\section{IV. The Fractal Peano basin}
Now we use our numerical tools to address the survival probability and the critical patch size for the fractal Peano basin. The Peano basin has a self-similar structure \cite{PeanoStructure1996,PeanoStructure1997} and its topological properties may be used to model dendritic like structures mimicking riverine ecological structure. Indeed, the connectivity of the environment, and in particular the river geometry, may affect the species extinction probability \cite{Fagan2002,Bertuzzo2011}.

The Peano network may be constructed from the following algorithm. For every new generation $Q$, any segment joining two sites is split, and three new sites are placed in the half of the segment. Figure \ref{fig4} represents this procedure. Thus, for a given generation $Q$, the total number of sites is $N=4^Q+1$. 

Now we present some general characteristics of the largest eigenvalue of $\mL$. We consider periodic boundary conditions along the backbone (that corresponds to the open circles in Fig. \eqref{fig4}). In the uniform environment ($\Delta=0$), we find numerically that $\lambda_1=-a$. Therefore, complete extinction occurs for $a>0$. In the presence of random growth rates ($\Delta \ne 0$), the survival probability, $P_s$, and the average of the largest eigenvalue, $\langle \lambda_1 \rangle$, exhibit the same qualitative features illustrated in Fig. \ref{fig1} for the linear case. In particular, we find that $p(\lambda_1)$ has a more pronounced asymmetry around $\langle \lambda_1 \rangle$ (as compared to the linear case), showing a small departure from the Gaussian shape. However, we still define the critical disorder strength according to $\langle \lambda_1\rangle(\Delta_c,N,D,a)=0$. In particular, we show that:

\begin{figure}
\includegraphics[width=0.7\columnwidth]{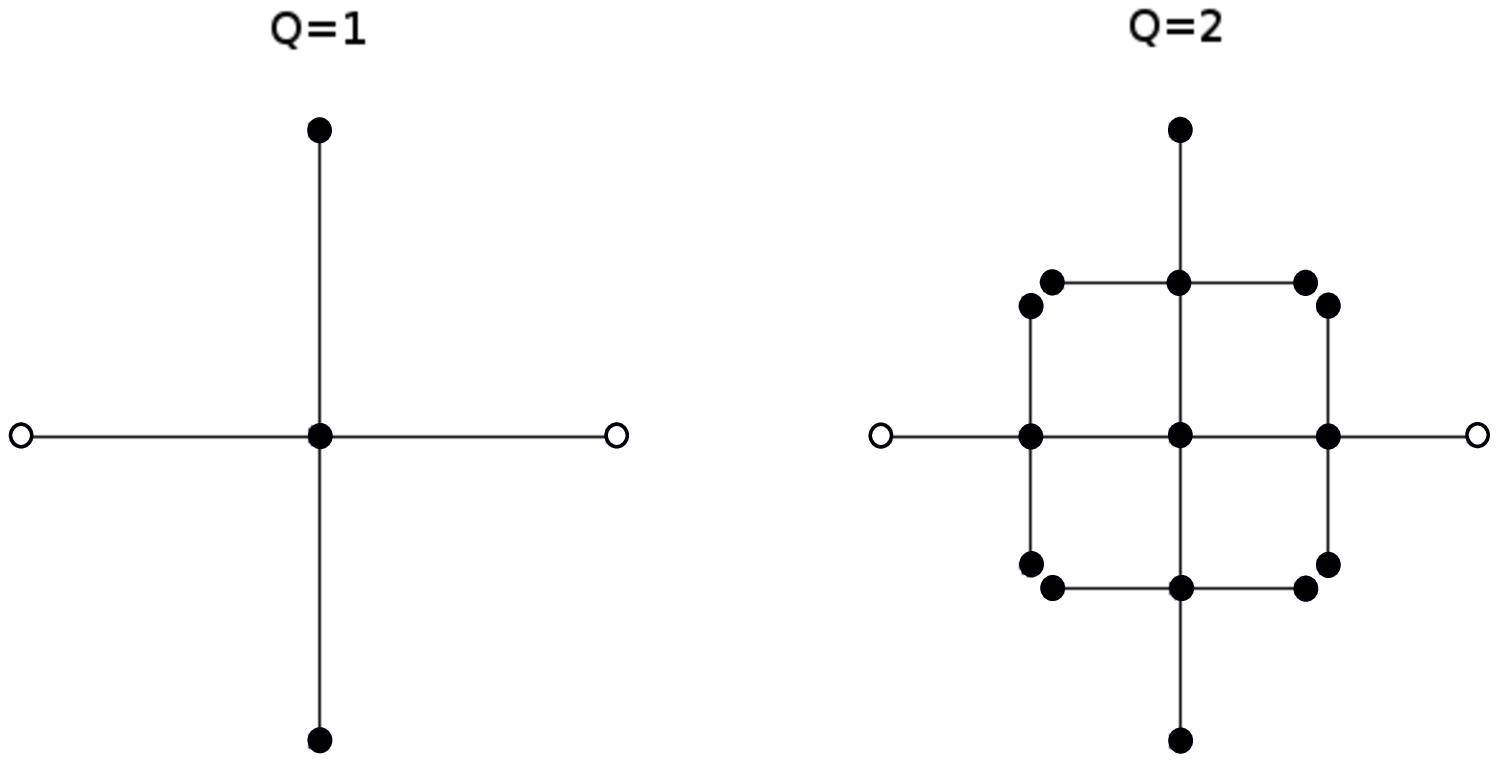}
\caption{Structure of the fractal Peano basin for the first ($Q=1$) and second ($Q=2$) generations.}
\label{fig4}
\end{figure}

\begin{equation}
\Delta_c(N,D,a)=c^{\prime}a^{1-\delta^{\prime}}(N^{-\phi^{\prime}}D)^{\delta^{\prime}}+a,\qquad (N\gg 1),
\label{powerlawPeano}
\end{equation}
where $\phi^{\prime}=1/4$ and $\delta^{\prime}=0.55\pm 0.01$ are the scaling exponents, and $c^{\prime}=1.27\pm 0.05$ is a constant. The actual values of $\delta^{\prime}$ and $c^{\prime}$ were obtained from the fitting analysis. The solid lines in the main box of Fig. \ref{fig5} (a) corresponds to expression \eqref{powerlawPeano} (the chi-squared of all the fits are less than $\sim 10^{-4}$). While the actual value of $\phi^{\prime}$ was obtained using finite-size scaling. In the inset of Fig \ref{fig5} (a) we show the collapse of $\Delta_c$ for $\phi^{\prime}=1/4$.

Now we address the critical patch size. We find that the eigenvector associated to $\lambda_1$ is spatially localized and its amplitude decays very fast across neighboring sites. Therefore, the participation ratio of the eigenvector associated to $\lambda_1$ still reflects the number of sites where the species are localized, and it may still be used to estimate $L_c$.

Unlike the one-dimensional case, the effect of the diffusion coefficient is drastic on the PDF of critical patch sizes. In Figure \eqref{fig5} (b) we study the evolution of $F(r_c)$ for increasing values of $D$. The first essential point is that the scaling-law \eqref{pL} is no longer valid for low diffusion regimes. However, it starts to be valid for higher values of $D$, when the mixing is so large that the topological structure is no more relevant.

In Fig. \ref{fig5} (c) we compare the participation ratio, $\langle r_c\rangle$, along the critical disorder strength (such that, $P_s\to 1/2$ for $\Delta\to \Delta_c$), between the Peano basin and the 1D case. As we might expect, the average value of $r_c$ for the Peano basin does not have the dependence on the square-root of the diffusion coefficient. We performed fits with a power law function, and we find that
\begin{equation}
\langle L_c \rangle/\ell_0 \approx \langle r_c \rangle = C^{\prime} \Big(\frac{D}{a} \Big)^{\beta^{\prime}} +1,
\end{equation}
with $\beta^{\prime}=3/4$ and $C^{\prime}=1.77\pm 0.03$. For low diffusion coefficient the size of the critical patch in the Peano basin is slightly smaller with respect to the one-dimensional case, i.e., the stationary population is more localized. However, the amount of positive fluctuations in the growth rate allowing for survival is always higher in the Peano case, as shown by comparison of Fig. \ref{fig2} (a) and Fig. \ref{fig5} (a). In other words, survival is always favored in the one-dimensional case.

The results we found on the scaling of $L_c$ suggest that the beta exponent depends on the fractal dimension of the system. In fact, our results are consistent with an exponent $\beta= d_w/4$ where $d_w$ is the exponent of the diffusion in the fractal $\sqrt{\langle r^2\rangle_t} \sim (Dt)^{1/d_w}$. In the d-dimensional case we have $d_w=2$ whereas in the Peano $d_w=3$ \cite{Lin}. For other fractals, according to our conjecture the beta exponent can be also irrational.

\begin{figure*}
\includegraphics[width=\textwidth]{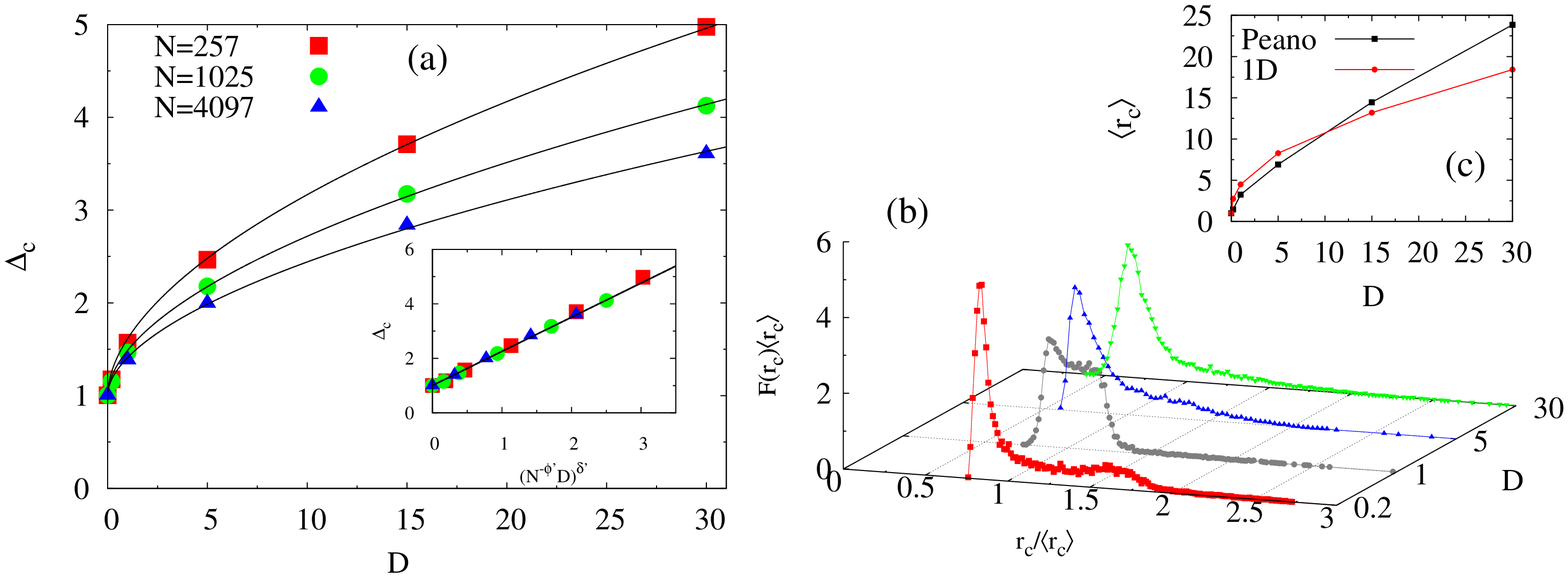}
\caption{Panel (a): Critical disorder strength $\Delta_c$ vs. $D$ for the fractal Peano basin. The number of sites considered are shown in the legend. The solid lines are fits according to expression \eqref{powerlawPeano}. The reduced chi-squared is $< 10^{-4}$ for all the fits. In the inset we linearize $\Delta_c$ using the exponent $\delta^{\prime}=0.55$ and show a quite good collapse using $\phi^{\prime}=1/4$. Panel (b): Scaling analysis of the PDF of participation ratios for various diffusion coefficients. Panel (c): We compare the average of the participation ratios between the fractal Peano basin ($N=1025$) and the 1D system ($N=1000$). The growth rate considered in all the plots is $a=1$.}
\label{fig5}
\end{figure*}

\section{V. Conclusions}
In this work we studied the population dynamics of a single-species in hostile and disordered environments for two different spatial network topologies: the 1D system and the fractal Peano basin. We firstly studied the conditions leading to extinction as well as survival of species, i.e., the survival probability. We then calculated the critical patch size needed to ensure survival of species in the long time dynamics, i.e., the critical patch size. We addressed these two problems by estimating both numerically and analytically the largest eigenvalue, $\lambda_1$, and the corresponding eigenvector, ${\bf v}_{\lambda_1}$, of the linearized Fisher dynamics.

From the probability distribution function (PDF) of $\lambda_1$ we were able to investigate the effect of different parameters (diffusion, size, etc.) on the critical disorder strength $\Delta_c$, which reflects a sort of critical extinction-survival transition (since that, $P_s\to1/2$ for $\Delta \to \Delta_c$). For both topologies the increase of the spatial fluctuations favor the species survival, and we provided explicit expressions for $\Delta_c$ (see Eqs. \eqref{powerlaw} and \eqref{powerlawPeano}). In particular, we have shown that the fractal Peano requires higher spatial fluctuations to ensure persistence, as compared to the 1D case, and using scaling arguments we find the corresponding scaling exponents, $\phi=1/3$ (1D) and $\phi^{\prime}=1/4$ (Peano).

Regarding the critical patch size, we demonstrated that the participation ratio, $r$, corresponding to ${\bf v}_{\lambda_1}$ can be used to estimate $L_c$, that is, $L_c\approx r$. This connection may be quite useful in situations where a numerical integration of Eq. \eqref{continuous_fisher} is highly costly, or in cases where an analytical approach to $L_c$ is challenging, for example, in a fractal topology. Furthermore, using this connection we were able to address the PDF of the critical patch sizes, $F(L_c)$. 

Our theoretical framework exploits concepts from Anderson localization and Random Matrix Theory to study and predict conditions of persistence and extinction of populations of replicating bacteria in a hostile environment, where only few, randomly distributed patches may support life. Similar ideas have been applied to study the biological evolution of simple organisms through the quasispecies model \cite{quasispecies1,quasispecies2,quasispecies3,quasispecies4}. The role of the spatial structure of the environment has an impact on the species localization and corresponding critical patch sizes, especially in low diffusion regimes. We thus found, as in other contexts \cite{Kerr2002,Bertuzzo2011}, the species survival may be favored, for a given fixed diffusion, by environments with lower average connectivity. A nature future direction will be to study, both theoretically and experimentally \cite{Giometto2016}, the role of spatial and temporal correlations of the environmental fluctuations on the species survival, and what is the spatial configuration of resources that may maximize the species stationary population.

\section{Acknowledgments}
R.P.R. thank Prof. Jos\'e A. Freire for useful comments, Prof. S\'ergio S. Rocha and Loren Kocillari for insightful discussions, and gratefully acknowledges the financial support from the Brazilian agencies CAPES (Grant number 12742/13-9) and CNPq (Grant number 201241/2015-3). W.F. acknowledges the Brazilian agency CNPq (Grant number 303253/2013-4) and INCT-FCX (FAPESP-CNPq 573560/2008-0).

\section{Appendix: Probability distribution function of critical patch sizes}
Using simple arguments we can obtain a fitting expression for $p(r)$. From Anderson localization we know that $v_{\lambda_1}(x)\sim \exp(-\kappa_0|x|)$, where $k_0$ is the inverse of the localization length (a positive quantity). Using this expression in Eq. \eqref{PR} we can show that $r\sim 2\kappa_0^{-1}$ ($N\to \infty$). The PDF of the critical patch sizes follows from the PDF of the inverse localization length \cite{Gardiner2003},
\begin{equation}
p(r)=\int_0^{\infty} \delta(r-2\kappa_0^{-1})g(\kappa_0)d\kappa_0.
\label{App_pdfPR}
\end{equation}
Based on the numerical results, we approximate $g(k_0)$ by a half-normal distribution, $g(\kappa_0)=A\exp\{-(\kappa_0-\mu)^2/2\sigma^2\}$ ($\kappa_0>0$), where $\mu$ is the mean, $\sigma$ is the variance and $A$ is a normalization constant. Using this expression in \eqref{App_pdfPR} we obtain
\begin{equation}
\label{App_fL}
p(r)= \frac{A}{(r)^2}\exp\Big\{-\frac{(\frac{2}{r}-\mu)^2}{2\sigma^2} \Big\}, \qquad r > 0,
\end{equation}
where $A$ is given by,
\begin{equation}
A=\frac{4}{\big(1+{\rm erf}(\frac{\sqrt{2}\mu}{2\sigma})\big)\sqrt{2\pi\sigma^2}}.
\end{equation}
The error function is defined as ${\rm erf}(x)= 2\pi^{-1/2}\int_0^x \exp(-t^2)dt$. Expression \eqref{App_fL} has two free parameters, and it can be used to fit the numerical data.

\end{document}